\documentclass[12pt]{article}
\usepackage{graphicx} % useful for figures
\newcommand{\half}{\frac{1}{2}}
\newcommand{\C}{{\cal C}}
\newcommand{\g}{{\bf g}}
\newcommand{\Zset}{{\relax{\sf Z\kern-.4em Z}}}
\newcommand{\Rset}{{\relax{\sf I\kern-.1em R}}}
\newcommand{\np}[3]{Nucl. Phys. {\bf B#1} (#2) #3}
\newcommand{\pl}[3]{Phys. Lett. {\bf #1B} (#2) #3}
\newcommand{\prd}[3]{Phys. Rev. {\bf D#1} (#2) #3}
\begin{document}

\begin{titlepage}

\begin{flushright}
{\tt UTTG-10-98}\\
{\tt UCLA-98-TEP-18}\\
{\tt hep-th/9806224}\\
\bigskip 
\end{flushright}
\bigskip
\null\vfil
\begin{center}
{\LARGE M-Theory Five-brane Wrapped on Curves for Exceptional Groups }
\vskip 3em
{\small \it 
     \lineskip .75em%
     \begin{tabular}[t]{ccc}
        \begin{tabular}[t]{c}\large \rm   Elena C\'aceres\\ \\
	Department of Physics,\\
	University of California at Los Angeles\\ 
	  Los Angeles, CA 90095-1547  USA\\
	{\rm e-mail:} caceres@physics.ucla.edu
	\end{tabular}
 &\rm\large  and &       
	\begin{tabular}[t]{c}\large \rm  Pirjo Pasanen\\ \\
	Theory Group, Department of Physics\\
	  University of Texas at Austin\\ 
	 Austin, TX 78712-1081 USA\\
	 {\rm e-mail:} ppasanen@physics.utexas.edu
 	\end{tabular}

     \end{tabular}
      \vskip 1.5em%
}
\end{center}
\vfil\null
\begin{abstract}

We study the M-theory five-brane wrapped around the Seiberg-Witten
curves for pure classical and exceptional groups given by an
integrable system. Generically, the
D4-branes arise as cuts that collapse to points after compactifying
the eleventh dimension and going to the semiclassical limit, producing
brane configurations of NS5- and D4-branes with $N=2$ gauge theories
on the world volume of the four-branes.  We study the symmetries of
the different curves to see how orientifold planes are related to the
involutions needed to obtain the distinguished Prym variety of the
curve. This explains the subtleties encountered for the
Sp($2n$) and SO($2n +1$).  Using this method we  investigate the
curves for exceptional groups, especially G$_2$ and E$_6$, and show
that unlike for classical groups taking the semiclassical ten
dimensional limit does not reduce the cuts to D4-branes.
For G$_2$ we find a genus two quotient curve that contains the Prym
and has the right properties to describe the G$_2$ field theory, but
the involutions are far more complicated than the ones for classical
groups. To realize them in M-theory instead of an orientifold plane we
would need another object, a kind of curved orientifold surface. 

\end{abstract}
\vfil\null
\end{titlepage}
%%%%%%%%%%%%%%%%%%%%%%%%%%%%%%%%%%%%%%%%%%%%%%%%%%%%%%%%%%%%%%%%%%%%%%%%%%%
\baselineskip 0.60cm
%\baselinestretch{2}

\section{Introduction}
\label{intro}

In the last years, D-brane techniques have proven useful in helping to
understand the strong coupling behavior of four dimensional gauge
theories with $N=2$ \cite{HanWit:3d,Wit:M} and $N=1$
\cite{Bar:Rot,Wit:BraQCD,EliGiKuRaSch} supersymmetries; for a review
and more comprehensive list of references see \cite{GivKut:Bra}.  The
$N=2$ supersymmetric Type IIA brane configurations consisting of NS
five-branes and D4-branes can be understood as coming from the
M-theory five-brane after compactification of one dimension
\cite{Wit:M}. The M5-brane world volume is given by $\Rset^4 \times
\Sigma$, where $\Sigma$ is the Seiberg-Witten curve holomorphically
embedded in $\Rset^4$.  N=2 theories with classical gauge groups and
matter in the fundamental, symmetric and antisymmetric representations
have been studied in this context
\cite{LanLopLow:Ori,BrSoThYa:SOSp,LanLop:New}.  Models with
exceptional gauge groups and more general matter representations have
not been discussed. Even though F-theory provides a framework where
exceptional groups can be investigated at strong string coupling
\cite{DaMu:F,Jo} it is not clear whether generalized Chan-Paton
factors can generate theories with D-branes and exceptional groups at
weak coupling \cite{GaZw:excep,GaHaZw:tensor1}.

Nevertheless, given that we know the Seiberg-Witten curves for
exceptional groups from the integrable systems
\cite{MarWar:Int,LerWar:E6} a natural question to ask is what happens
if we wrap the M-theory brane on such a curve.  The purpose of this
paper is to investigate what those configurations will reproduce after
compactification to ten dimensions. And if the brane description of
$N=2$ gauge theories breaks down for the exceptional groups, to
understand why.

Furthermore, finding a brane construction for SO groups with spinors
would be specially interesting since Seiberg dual pairs containing
spinors are known \cite{Poul:dual,BeChoKrStr:so10} and they should
correspond to some embedding in string theory.  A natural way to
combine these issues would be to investigate E$_6$ and its breaking to
SO(10).  Thus it would be important to understand gauge theories
with exceptional groups within the brane context.

We start from a Seiberg-Witten curve for pure gauge groups, known from
integrable Toda systems \cite{MarWar:Int} and wrap a M5-brane around
it. We investigate what parts of the curve will produce the NS5- and
D4-branes in the ten dimensional limit.  We obtain a well defined
procedure for constructing brane diagrams, explaining some subtleties
in the literature encountered especially for symplectic and odd
orthogonal groups. Using these methods we can try to apply them to
exceptional groups.  We discuss how the configuration obtained from an
exceptional curve after compactifying to ten dimensions is
fundamentally different from the ones for classical groups and why it
does not seem to give a world volume gauge theory. The reasons boil
down to the fact that the branch cuts, which for classical groups will
reduce to D4-branes in ten dimensions, will not do so for the
exceptional groups.  Also, for the unitary, orthogonal and symplectic
groups the  curve of genus = rank$(G)$ whose Jacobian is the distinguished
Prym variety can easily be obtained as a quotient  by $\Zset_2$
involutions. These involutions can be naturally realized in the
ten dimensional brane picture as projections by orientifold planes.
For exceptional groups this is not possible.

We will focus on G$_2$ because it allows us to understand certain
features unique to exceptional groups without being overly
complicated.  We are able to find the genus two subvariety of the
G$_2$ curve that contains the Prym, but the symmetries involved cannot
be described as orientifold planes in ten dimensions. Instead we find
that we would need a new kind of object in M-theory, which seems to be
a curved orientifold surface.

After reviewing some preliminaries we show in section \ref{brabra} the
relation between D4-branes and branch cuts of the Seiberg-Witten
curves.  In section \ref{SymmPrym} we investigate the role played by
the symmetries of the curve and how they will help to determine the
brane configurations in ten dimensions.  In section \ref{G2} we apply
the previous results to the G$_2$ and E$_6$ theories and state our
conclusions in section \ref{conclusions}.

%%%%%%%%%%%%%%%%%%%%%%%%%%%%%%%%%%%%%%%%%%%%%%%%%%%%%%%%%%%%%%%%

\section{Preliminaries}
\label{prelim}

Four dimensional N=2 gauge theories arise in Type IIA context when we
consider a five-brane with a world volume given by $\Rset^4 \times
\Sigma$, where $\Sigma$ is the Seiberg-Witten Riemann surface embedded
in a four dimensional space \cite{KlLeMaVaWa}. This picture can be
reinterpreted in M-theory \cite{Wit:M} by considering an eleventh
dimension $x^{10}$, taken to be periodic: $x^{10}\sim x^{10} + 2 \pi
R$. The Riemann surface is now embedded in $\Rset^3\times S^1$. 
Taking the radius $R$  to zero, i.e. going to ten dimensions, 
the M-theory five-brane produces  a Type IIA
configuration of NS5- and D4-branes.  The NS5-brane is the M5-brane on
$\Rset^{10}\times S^1$, whose world volume is located at a point in
$S^1$ and thus spans a six dimensional manifold in $\Rset^{10}$.  The
D4-brane is the M5-brane wrapped over $S^1$; its world volume projects
to a five dimensional manifold.

The ten-dimensional brane configurations consist of D4-branes
stretched in between NS5-branes and possibly some D6-branes and
orientifold four- and six-planes, depending on the Riemann surface
chosen.  Following the standard conventions we will consider NS
five-branes with world volume along $x^0,x^1....x^5$ and located at
$x^7=x^8=x^9=0$ and at an arbitrary value of $x^6$.  The D4-branes
O4-planes extend along $x^0,x^1,x^2,x^3,x^6$ but the four-branes are
finite (of length $L_6$) in the $x^6$ direction.  They live
(classically) at a point in $x^4,x^5$ and they are located at
$x^7=x^8=x^9=0$. Orientifold six-planes and D6-branes extend along
$x^0,x^1,x^2,x^3,x^7,x^8,x^9$.

$N=2$ supersymmetry demands that the four-dimensional space
$\Rset^3\times S^1$ where the Riemann surface $\Sigma$ lives on should be
complex. Moreover, $\Sigma$ has to be holomorphically embedded in it
with respect to coordinates $v$ and $t$, defined as
\begin{eqnarray}
v&=& x^4 +{\mathrm i} x^5\nonumber\\
s&=&(x^6 +{\mathrm i} x^{10})/R\nonumber\\
 t&=&\exp(-s).
\label{defvst}
\end{eqnarray}
The vector multiplets of the D4-brane world volume field theory
originate from the chiral antisymmetric tensor field living on the
M-theory five brane. If $\Sigma$ is a compact\footnote{In the case of
five branes and four branes suspended between them the curve $\Sigma$
is actually not compact, but can be compactified by adding points, see
\cite{Wit:M}.} Riemann surface of genus $n$ the zero-modes of the
antisymmetric tensor give $n$ abelian gauge fields. The low energy
effective action of these fields is determined by the Seiberg-Witten
differential $\lambda_{SW}$ and $\alpha$ and $\beta$-cycles of
$\Sigma$. In particular, the couplings of the gauge fields are
determined by the periods $\int_\alpha \lambda_{SW}$ and $\int_\beta
\lambda_{SW}$, i.e. the Jacobian $J(\Sigma)$ of the Riemann
surface.

There is a correspondence between Seiberg-Witten curves and integrable
systems
\cite{ItoMor:intSW,GKMMM,MarWar:Int,MarWar:proof,DHokPho:CalMos,
DonWit:YMInt}.  In \cite{MarWar:Int} Martinec and Warner showed that
the relevant curve for a $N=2$ pure SYM theory with gauge group $G$
arises as the spectral curve of a periodic Toda lattice for the dual
affine Lie algebra $\g^\vee$ of the group $G$.  In a Toda system the
powers $1/t$ of the spectral parameter appear at a grade related to
the Coxeter number of the group, $h_\g$. Martinec and Warner observed
that in the $N=2$ SYM theories the instanton generated term $\mu/t$
appears at a grade $2h_\g^\vee$, $h_\g^\vee$ being the dual Coxeter
number.  For non-simply laced groups $h_\g^\vee \neq {h_\g}$ and thus
we should relate $N=2$ Yang Mills theories to the Toda curve for the
corresponding dual affine algebra. For simply laced groups the
distinction is irrelevant. The dual Coxeter numbers are listed in the
following table:
\vskip 1em
\begin{tabular}{|l|c|c|c|c|c|c|}       \hline
group $G$ & SU($n$) & SO($2n$) & Sp($2n$) & SO($2n+1$)& G$_2$& E$_6$
\\ \hline
$h_\g^\vee$ & $n+1$ & $2n-1$& $n+1$ &$2n-2$ &  4&  12 \\ \hline
\end{tabular}
\vskip 1em
In gauge theory the $\mu$ parameter sets the quantum scale:
 $\mu \sim {\Lambda}^{2{h_\g}^\vee}$,
where $\Lambda$ is the energy scale of the theory.

The curves obtained in \cite{MarWar:Int} and \cite{LerWar:E6} 
are\footnote{Note that
there is a mistake in the original paper for the Sp($2n$). The curve
listed here is the one obtained from the Lax matrix for the
 twisted affine algebra of Sp($2n$).}:
\begin{equation}
\begin{array}{rl}
{\rm SU}(n):& 	   t+\mu/t -(v^{n}+u_2 v^{n-2} +\cdots +u_{n})=0\nonumber\\
{\rm SO}(2n):&   	    v^2(t+\mu/t) -(v^{2n} +u_2 v^{2(n-1)}+ \cdots 
			+u_{2n-2} v^2 +u_{2n}) =0 \nonumber\\
{\rm SO}(2n+1):& v(t+\mu/t) -(v^{2n} +u_2 v^{2(n-1)} +
			\cdots + u_{2n})=0\nonumber\\
{\rm Sp}(2n):&   (t+\mu/t)^2 - v^2(v^{2n} +u_2 v^{2(n-1)} +\cdots
			 +u_{2n})=0 \label{curves} \\
{\rm G}_2:&     3(t-\mu/t)^2 +2(t+\mu/t)[u_2v^2-3v^4] -v^8 +2u_2v^6- u_2^2 v^4+
  u_6 v^2=0\nonumber\\
{\rm E}_6:& x^3(t+\mu/t -u_{12})^2 - 2(t+\mu/t -u_{12}) q_{15}(v) -
 \frac{1}{ x^3}[q_{15}^2(v) - p_{10}^2(v) r_{10}(v)],\nonumber
\end{array}
\end{equation}
where $u_i$ denotes the $i$:th order invariant of the group and
$q_{15}(v), p_{10}(v) $ and $r_{10}(v)$ are polynomials of degrees 15,
10 and 10, respectively.

%%%%%%%%%%%%%%%%%%%%%%%%%%%%%%%%%%%%%%%%%%%%%%%%%%%%%%%%%%%%%%%%%%

\section{Branes and Branch Cuts}
\label{brabra}

We want to identify what parts of the curves will produce the NS5- and
D4-branes when we compactify to ten dimensions. To be more precise, we
will also need to take a semiclassical limit to obtain the usual brane
configurations.  Recall that $\mu \sim \Lambda^{2{h_\g}^\vee}$. Thus
from the four dimensional gauge theory point of view the classical
limit $\Lambda \rightarrow 0$ implies taking $\mu \rightarrow 0$.  On
the other hand, in string theory the classical limit is obtained by
letting the gauge coupling $g \rightarrow 0$ at the same time as
$g_{s} \rightarrow 0$ and $L_6/l_s \rightarrow 0$. Since the radius of
the eleventh dimension is $R_{10}= g_{s} l_{s}$ the semiclassical
limit implies also taking $R_{10}\rightarrow 0$. 

To see where the Type IIA branes originate from we consider 
a curve $\Sigma$ defined by  equation $F(t,v)=0$.
For simplicity we will  first take $F$ to be of second order in $t$
\begin{equation}
F(t,v)= A(v) t^2 +B(v) t +C(v) =0
\label{doubcov}
\end{equation}
It was argued in \cite{Wit:M} that this curve represents two
NS5-branes with $k$ D4-branes suspended in between them ($k$ being the
degree of $B(v)$) and that the degrees of $A(v)$ and $C(v)$ give the
number of semi-infinite D4-branes extending to the left and to the
right of the leftmost and rightmost five brane respectively.
Following this approach different suggestions for brane configurations
of SO($n$) and Sp($2n$) groups were made
\cite{BrSoThYa:SOSp,LanLopLow:Ori}. While this approach works well for
SU($n$) there are some subtleties in the case of other gauge
groups. We will show that from the M-theory point of view the
D4-branes correspond to the branch cuts of $\Sigma$ viewed as a double
cover (or $n$-fold cover if $F(t,v)$ is of degree $n $ in $t$) of the
$v$-plane.  The two sheets of the double cover form the two
NS5-branes, connected by the branch cuts.  In the $R_{10}\rightarrow
0$ limit this coincides with Witten's description of the classical
positions of the D4-branes.  The subtleties encountered in the
constructions of brane configurations for SO($n$) and Sp($2n$) groups
\cite{LanLopLow:Ori,BrSoThYa:SOSp,AhnOhTatar:sp,AhnOhTatar:so,AhnOhTatar:o6}
can be easily explained in this way.

\begin{figure}
  \noindent
  \begin{minipage}[b][11cm]{.46\linewidth}
    \centering\includegraphics[width=0.75\linewidth]{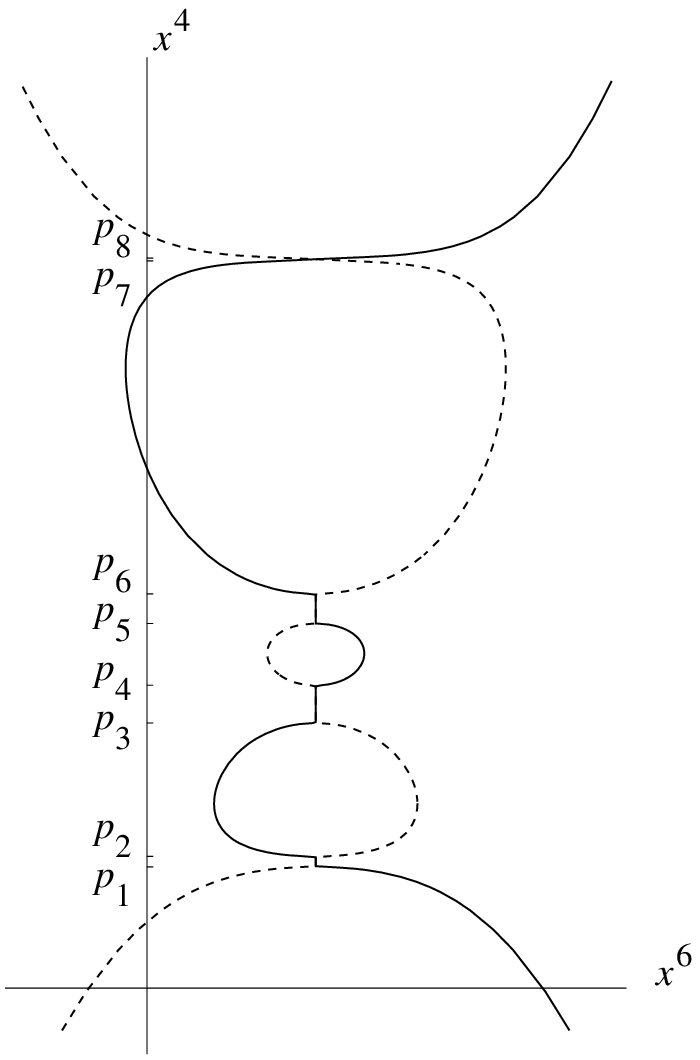}
    \caption{SU(4) real part. 
 The moduli have been chosen so that all the branch points $p_i$ are
located in the real $v$-axis $x^4$. 
The different types of lines represent the two sheets.}\label{SU4fig}
  \end{minipage}\hfill
  \begin{minipage}[b][11cm]{.46\linewidth}
    \centering\includegraphics[width=0.75\linewidth]{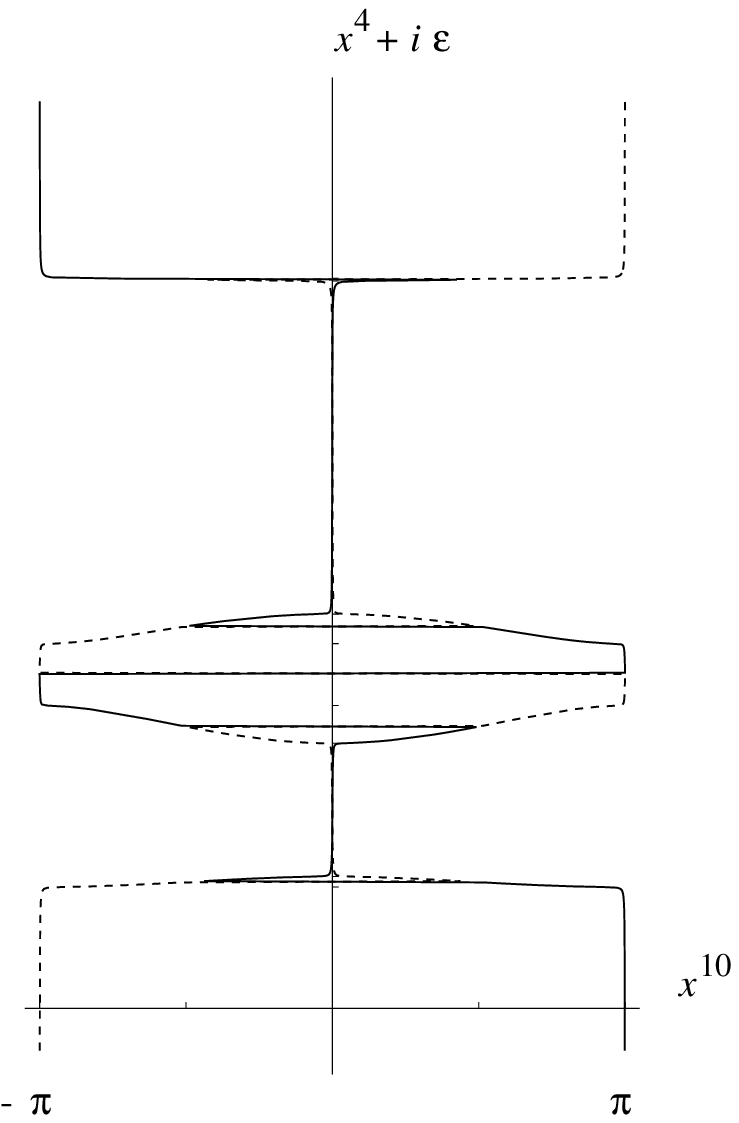}
    \caption{SU(4) imaginary part.
The point in  between $p_4$ and $p_5$ where $x^{10}$ jumps
$2 \pi$ is not a branch point but an artifact of the periodicity of
  $x^{10}$. For clarity, the plot is taken at small fixed 
imaginary part for $v$.}
\label{SU4imfig}
  \end{minipage}
\end{figure}

\begin{figure}
  \noindent
  \begin{minipage}[b][11cm]{.46\linewidth}
    \centering\includegraphics[width=0.75\linewidth]{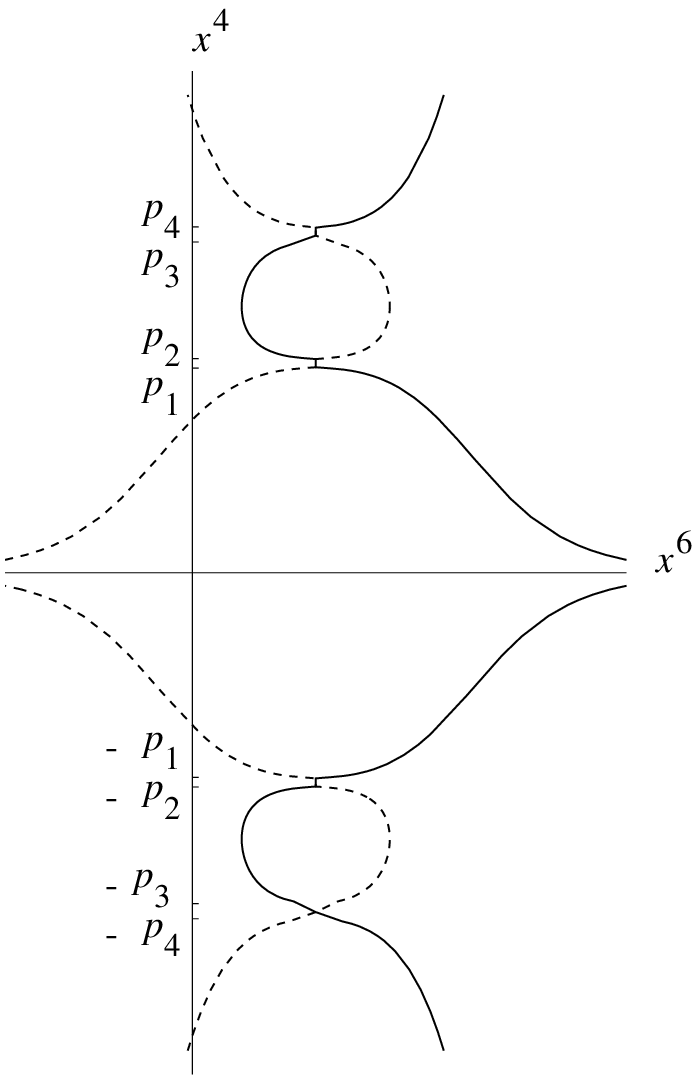}
    \caption{SO(4) real part.}\label{SO4fig}
  \end{minipage}\hfill
  \begin{minipage}[b][11cm]{.46\linewidth}
    \centering\includegraphics[width=0.75\linewidth]{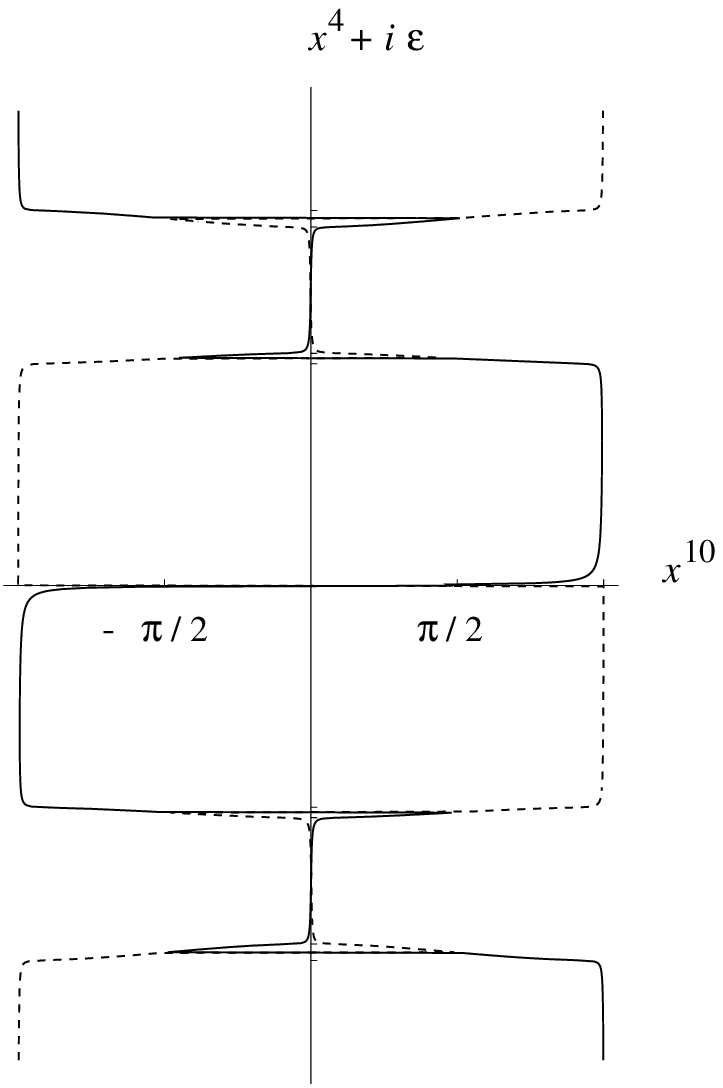}
   \caption{SO(4) imaginary part.}
  \end{minipage}
\end{figure}

\begin{figure}
  \noindent
  \begin{minipage}[b][11cm]{.46\linewidth}
    \centering\includegraphics[width=0.75\linewidth]{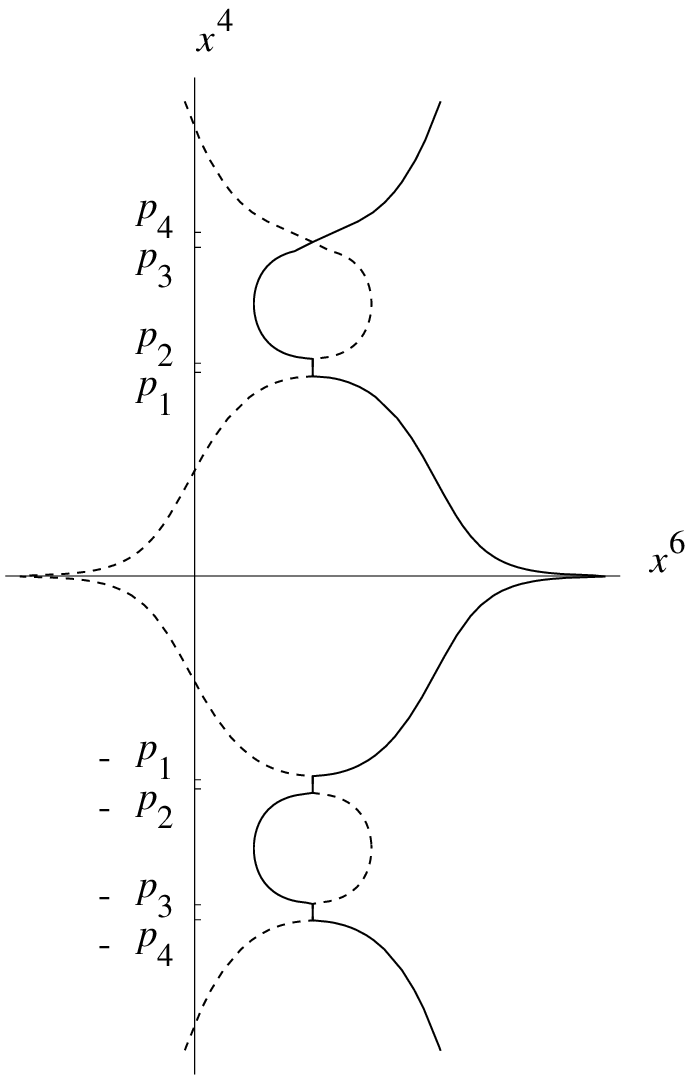}
    \caption{SO(5) with same values for the moduli as for SO(4) 
in Fig.\ref{SO4fig}}\label{SO5fig}
  \end{minipage}\hfill
  \begin{minipage}[b][11cm]{.46\linewidth}
    \centering\includegraphics[width=0.75\linewidth]{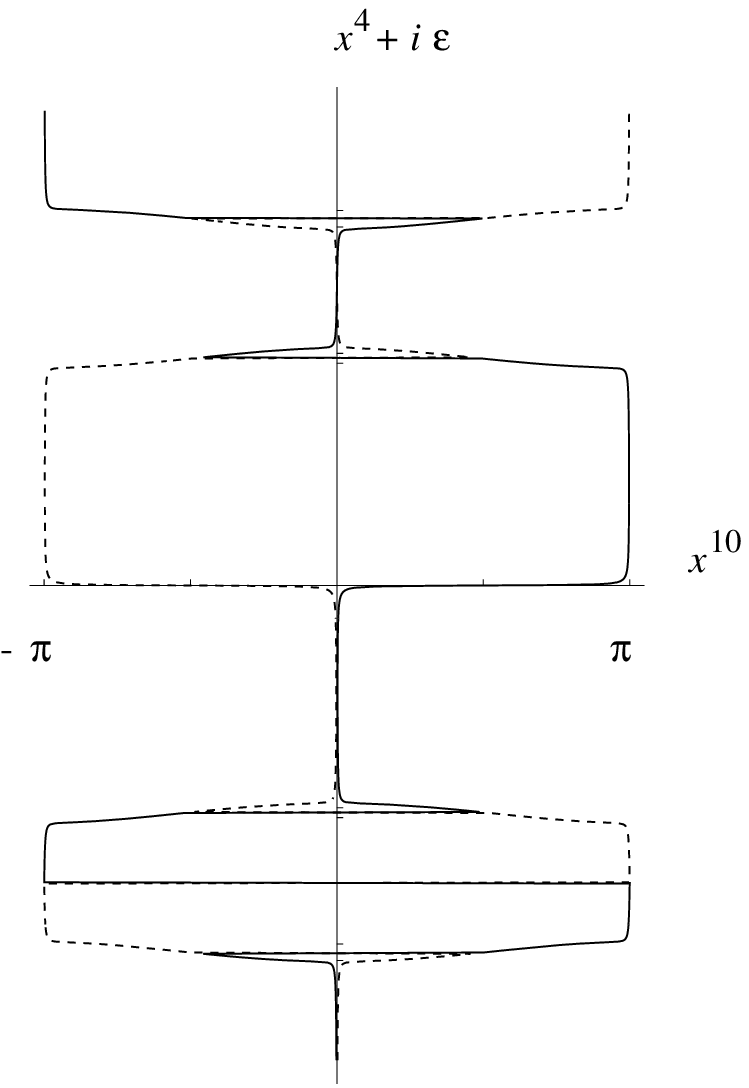}
   \caption{SO(5) imaginary part.}
  \end{minipage}
\end{figure}

\begin{figure}
  \noindent
  \begin{minipage}[b][11cm]{.46\linewidth}
    \centering\includegraphics[width=0.75\linewidth]{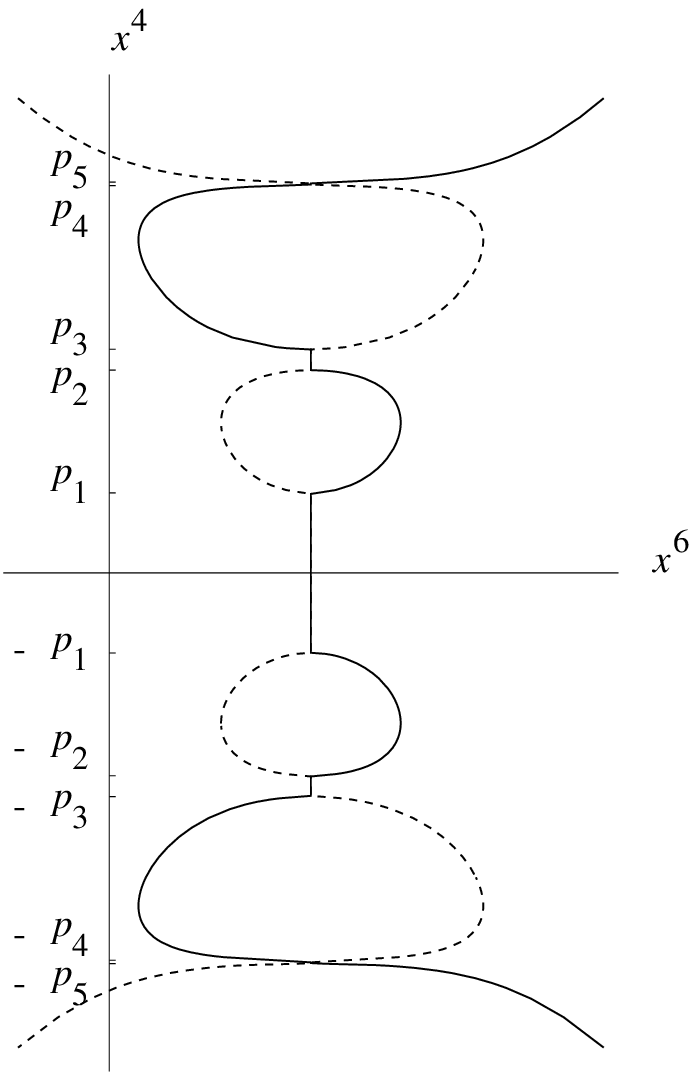}
    \caption{Sp(4) real part. This is the same for both the two-fold
  and four-fold cover curves.}
    \label{Sp4fig}
  \end{minipage}\hfill
  \begin{minipage}[b][11cm]{.46\linewidth}
    \centering\includegraphics[width=0.75\linewidth]{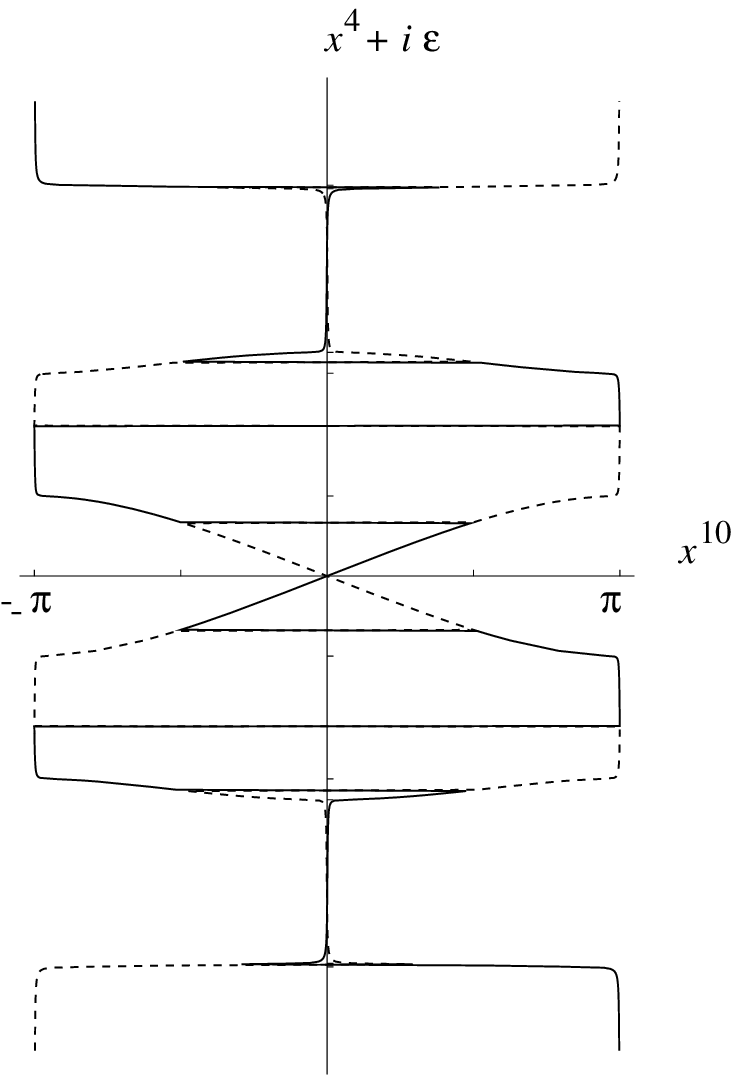}
   \caption{Sp(4) imaginary part for the two-fold cover curve.}
	\label{Sp4imfig}
  \end{minipage}
\end{figure}

By solving for $s$ in (\ref{doubcov})
\begin{equation}
s = -R \log \left( -\half B(v) \pm \half\sqrt{B(v)^2 - 4 A(v) C(v)}
\, \right)
\label{s-doubcov}
\end{equation}
we see that the branch points of $s$ are located on the $v$-plane at
$\Delta=B(v)^2 -4 A(v) C(v)=0$.  If we go around any of the branch
points we will find a discontinuity in the phase of $t$ (thus in
$x^{10}$) whenever we cross a branch cut.  Therefore, it is the cuts
in the $v$ plane that produce the wrapping around the $x^{10}$
direction. This can be seen explicitly for example in Figures
\ref{SU4fig} ans \ref{SU4imfig}
where we plot the real and imaginary parts of $s$ for
SU(4), with $A(v)=1,\ C(v)=\mu$ and $B(v) = P_{4}(v;u_i)$. 
For examples of other groups see Figures \ref{SO4fig}--\ref{Sp4imfig}.

We can now identify the number and position of the cuts of $\Sigma$
with the number and positions of the D4-branes.  More precisely, it is
the cuts in the semiclassical, ten dimensional limit $\mu \rightarrow
0$, $R_{10}\rightarrow 0$ that are the D4-branes. Therefore, a
requirement to have well defined D4-branes in ten dimensions is that
each branch cut collapses to a point.  Again using SU($n$) as an
example, we see that in the $\mu \rightarrow 0$ limit the discriminant
$\Delta=B_{n+1}^2(v) -4 \mu$ becomes a perfect square, each pair of
branch points degenerates to a point, the branch cut disappears and we
recover Witten's description of the position of the D4-branes as the
zeroes of $B_{n+1}(v)$.  The same thing happens for all other
classical groups as well. This can be seen by examining their
$\Delta$'s listed in the following table:
\vskip 1em
\begin{tabular}{|l|c|c|c|}       \hline
group $G$ &$\Delta $    &\# cuts = $\half$\# b.p. &genus of $\Sigma$\\ 
\hline
SU($n+1$)  &$B_{n+1}(v)^2 -4 \mu$    &$n+1$     &$ n$\\ \hline
SO($2n$)   &$B_n(v^2)^2 -4 \mu v^4$  &$2n$ 	&$2n-1$\\ \hline
SO($2n+1$)&$B_n(v^2)^2 - 4 \mu v^2$ &$2n $ &$ 2n-1$\\ \hline
Sp($2n$)   &$B_n(v^2)(v^2 B_n(v^2) -2\mu)$  &$2n+1$ &$2n$\\ \hline
\end{tabular}
\vskip 1em
\noindent
However, as we will see in section \ref{G2} the behaviour of the
exceptional curves in this limit is very different.

\subsection{Multiple NS5-branes and product gauge groups}

When $F(t,v)$ is of second order in $t$ identifying how the sheets
combine to form NS5-branes was rather straightforward. The only
subtlety being that the five branes should not ``cross'' each other in
the $x^6$-direction at the branch points to avoid confusion on which
brane is on the left and which on the right.

\begin{figure}
  \noindent
    \centering\includegraphics[width=6cm]{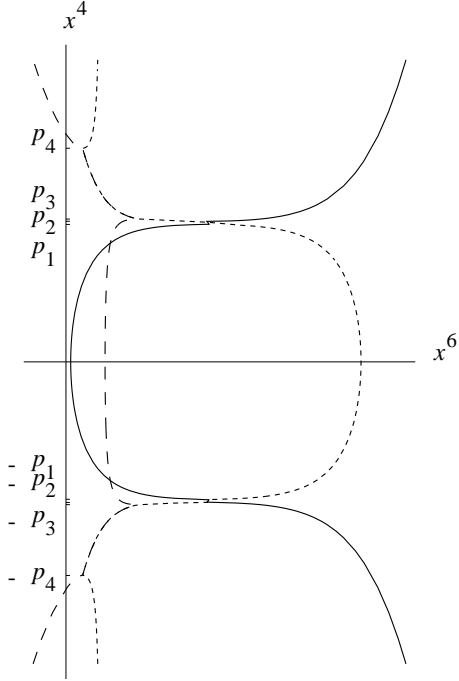}
    \caption{SU(2)$\times $SU(2) with bi-fundamental matter}
\label{prod-fig}
\end{figure}

For higher number of NS5-branes the situation is more complicated: we
can always determine the number of sheets and number of cuts, but
trying to identify which sheets are connected by which cuts depends
crucially on how we choose the phases at each cut. Take for example
the three-fold cover curve that describes the product group ${\rm
SU}(n)\times {\rm SU}(m)$ with bi-fundamental matter (see
Fig. \ref{prod-fig} for $n=m=2$):
\[
t^3 + t^2 \left( B_n(v) + \frac{\mu_m}{\mu_n}\right) +  t\left( B_m(v)
\frac{\mu_m}{\mu_n} + \mu_n\right) +   \frac{\mu_m^2}{\mu_n} =0,
\] 
here  $\mu_j$ is the scale of SU($j$) gauge
group.

The branch points are still of second order (they connect only two
sheets), but choosing the phases in a globally consistent way can be
somewhat complicated. However, there is an additional piece of
information we can use, namely the asymptotic behaviour of the sheets
as $v\to\infty$. Looking at the curve for large, small and middle
values of $t$ we see that they behave as $\ln v^2,\ \ln v^{-2}$ and
constant at large $v$. This means that the effective number of
four-branes attached to each five-brane should be $2, -2$ and 0,
respectively, where D4-branes on the left count as positive and on the
right as negative.  The classical limit now corresponds to taking
$\mu_n, \mu_m \to 0 $ separately for each group. In other words we
first decouple for example the second factor by $ \frac{\mu_n}{\mu_m}
\to \infty$, giving $\Delta \sim B_n(v)^2 - 4 \mu_n$, and then take
$\mu_n\to 0 $ to localize the D4-branes at the zeroes of
$B_n(v)$. Similarly for the SU($m$) factor. This reduces the curve to
the usual ten dimensional brane configuration of a product of groups,
studied for example in \cite{ErlNaqRan:Prod}.

Before examining other gauge groups in detail we will discuss the role
of the symmetries of the curves and how they will translate to
orientifolds in the brane pictures.

%%%%%%%%%%%%%%%%%%%%%%%%%%%%%%%%%%%%%%%%%%%%%%%%%%%%%%%%%%%
\section{Symmetries and Orientifolds}
\label{SymmPrym}

In most cases the curves given by the Toda-system have genus much
higher than the rank of the group --- only for SU($n+1$) with
Lax-matrix in the fundamental representation do they agree.  So in
addition to finding the differential $\lambda_{SW}$, one needs to
choose rank$(G)$ $\alpha$ and $\beta$-cycles and thus pick out the
physically relevant rank$(G)$-dimensional subvariety of the Jacobian
\cite{MarWar:Int}.  This sub-variety is called the preferred Prym. It
is the same for all curves that correspond to the same gauge group,
regardless of the representation of the Lax-matrix.  Also, the curves
have natural Weyl group actions acting on them. This induces
symmetries on the cycles and differentials and therefore on the
Jacobian. The preferred Prym variety is then the part of the Jacobian
which corresponds to the reflection representation of the Weyl group.

For most gauge groups it is possible to find explicitly a genus $n$
curve for which the Jacobian is just the Prym variety. These curves
are obtained from the original ones as quotients by certain
symmetries, see for example \cite{Don:SW} for explicit
constructions. The physics of the $N=2$ SW gauge theory for the
quotient curve is the same as the for the original curve.

It is natural to assume that the ten-dimensional brane picture should
inherit the symmetries of the curve we wrap the M5-brane on.  Because
the symmetries originate from the Weyl group action we should see that
action on the brane diagram.  For SU($n+1$) this does not give us
anything new: the curve already has genus $n$ and the Weyl group acts
by permuting the D4-branes. But for other gauge groups we can use this
to determine what the brane configuration should be.

The curves of orthogonal and symplectic groups have only $\Zset_2$
symmetries. The SO($2n$) and Sp($2n$) curves (\ref{curves}) are both
invariant under the involutions
\begin{eqnarray} 
\sigma_1 &:& \quad v\to -v \label{sigma12}\\
\sigma_2 &:& \quad t\to \mu /t\nonumber .
\end{eqnarray}  
Sp($2n$) has also an additional symmetry 
\begin{equation}
\sigma_3: \quad t\to -t. \nonumber \label{sigma3}
\end{equation}
 SO($2n+1$) curve on the other hand is left
invariant only under $\sigma_2$ and the combination $\sigma_1\sigma_3$.
If we rescale  $t$ by
\begin{equation} 
t= \sqrt{\mu} \exp [-(x^6 + {\mathrm i} x^{10})/R\, ]
\end{equation}
we see that $\sigma_2$ corresponds to $x^6\to -x^6,\ x^{10}\to
-x^{10}$ and $\sigma_3$ to a shift in $x^{10}$ coordinate: $x^{10}\to
x^{10}+ {\mathrm i}\pi R$.  These symmetries can be implemented to the ten
dimensional brane diagrams by adding an orientifold
\cite{EvaJonSha}. There are two possibilities: either an O4-plane
acting on the coordinates as
\begin{eqnarray}
O4:\quad (x^4,x^5, x^7, x^8,x^9, x^{10}) \to (-x^4,-x^5, -x^7,
-x^8,-x^9, -x^{10})
\end{eqnarray}
 or an O6-plane
\begin{equation}
O6:\quad (x^4,x^5,x^6, x^{10})\to (-x^4,-x^5,-x^6, -x^{10}).
\end{equation}
The action of the O6 in the $v$ and $s$ coordinates is exactly the
combined symmetry $\sigma_1\sigma_2$, whereas the O4 acts only as
$\sigma_1$.  With only two NS5-branes there is not much difference
whether we will us an O4 or an O6, but with more then two NS5-branes
there are problems associated to the charge of the O4 --- it should
change sign whenever the orientifold passes a five-brane
\cite{EvaJonSha}.  Moreover, if we require that the orientifold
projection should be equivalent to taking the quotient with respect to
all the symmetries of the curve we find that the O6 is more natural.

\subsection{Orientifold planes  in M-theory}

The reason why the O6 symmetries are related to the symmetries of the
curves becomes more clear when we look at their eleven dimensional
origin --- they are certain types of singularities in M-theory
\cite{Seib:IRDyn,SeiWit:Comp3d,Sen}.  Away from the singularity the
complex structure of these spaces can be given by
\begin{equation}
xy=\mu v^{2k},
\label{AH-space}
\end{equation}
where $2k$ is the RR-charge of the orientifold and $\mu$ the parameter
that sets the gauge theory scale: $\mu \sim \Lambda^{2h^\vee_\g}$;
$x,y\sim \Lambda^{2h^\vee_\g+k} $. When the charge $2k$ is negative
(\ref{AH-space}) describes a complex structure of the Atiyah-Hitchin
space. The interpretation for the positive charge is less
clear. Recall that adding to the gauge theory hypermultiplets in the
fundamental representation corresponds to modifying the singularity
structure by introducing $D6$-branes:
\begin{equation}
xy=\Lambda^{2h^\vee_\g}v^{2k-N_f}\prod_i^{N_f}(v-m_i).
\end{equation}
Thus the positively charged O6 resembles  $k$ D6-branes stuck in the
origin, but with a singularity that can not be resolved.

The gauge theory curves of classical groups can now be described
by the following equations:
\begin{equation}
\begin{array}{rll} 
 {\rm SU}(n+1):&  x+y - P_{n+1}(v)=0, &  2k= 0 \\
 {\rm SO}(2n):	 &  x+y - P_{2n}(v^2)=0, &  2k=4 \\
 {\rm SO}(2n+1):&  x+y - P_{2n}(v^2) =0,  &  2k=2  \\
 {\rm Sp}(2n):	 &  (x+y)^2 - P_{2n}(v^2)=0, &  2k=-2 \\
 \end{array}
\label{M-curves}
\end{equation}
After using (\ref{AH-space}) to solve for $x$ and defining $y= t
v^{k}$ we get the curves listed in (\ref{curves}). Note that here the
SO($2n+1$) curve is symmetric under $v\to -v$, only the scaling of $y$
changes that. Also, the charge of the orientifold is not absolutely
determined. For SO($2n+1$) we could as well have written $2k=4$ and
$x+y - v P_{2n}(v^2) =0$.

%%%%%%%%%%%%%%%%%%%%%%%%%%%%%%%%%%%%%%%%%%%%%%%%%%%%%%%%%%%%%%%%
\section{Orthogonal and Symplectic Groups}
\label{OSp}

The brane diagram for SO($2n$) is easily determined. As we have seen,
the symmetries of the curve are exactly those of O6. The ten
dimensional configuration of branes consists of two NS5-branes, an
orientifold six-plane in the origin and $2n$ D4-branes symmetrically
on each side of $x^6$-axis. In addition, we see from Figure
\ref{SO4fig} that the NS5-branes extend to infinity in the $x^4$ due
to bending caused by the orientifold. Those can be interpreted as two
semi-infinite D4-branes on each side, located at $x^4=0$.  The
O6-plane effectively mods out all the symmetries, leaving only the
physically significant part. In the original curve \( \C_{{\rm
SO}(2n)}: v^2(t+ \mu /t) + P_{2n}(v^2) =0 \) this amounts to defining
new variables $u=v(t-\mu /t)$ and $\xi=v^2$, giving the genus $n$
surface where the Prym lives:
\begin{equation}
\C'_{{\rm SO}(2n)}:\quad\xi u^2 = P_n(\xi)^2 - 4 \mu\xi^2 .
\end{equation}
Note that we can not wrap the M-theory five-brane directly on this
reduced curve and still get the same world volume gauge theory.  The
reason is that to get the correct gauge group we need the orientifold
to project out part of the Chan-Paton factors.
 
The SO($2n+1$) is a bit more complicated. As mentioned in the previous
section, there are two equivalent descriptions: one with O6 of charge
$2k=2$ and $\C_{{\rm SO}(2n+1)}: v(t+ \mu /t) - P_{2n}(v^2) =0$ and
the other with a charge 4 orientifold and slightly different
polynomial \(\C_{{\rm SO}(2n+1)}: v^2(t+ \mu /t) - v P_{2n}(v^2) =0\).
The former would describe a brane diagram with one semi-infinite
D4-brane on each side of the five-branes at $v=0$ and nothing between
them.  The latter has also an extra D4-brane sitting on top of the
orientifold, extending all the way to plus and minus infinity. So
effectively this is the same as the SO($2n$) configuration but with
one D4 in the origin $v=0$, between the five-branes, making it a total
of $2n+1$ D4-branes. This is the configuration suggested also in
\cite{BrSoThYa:SOSp}.  In both cases, as for SO($2n$), the
semi-infinite branes do not represent matter but are a result of the
bending of the M5-brane in the presence of the orientifold.  The genus
$n$ quotient curve is very similar to that of SO($2n$):
\begin{equation}
\C'_{{\rm SO}(2n+1)}:\quad\xi y^2 = P_n(\xi)^2 - 4\mu \xi ,
\end{equation}
where $y=t-\mu/t$ and $\xi=v^2$.

The Sp($2n$) curve differs from the curves of all other classical
groups in that it is of fourth order in $t$, meaning that we should
have a brane configuration with four NS5-branes. However, looking at
the symmetries of and restricting to the physically significant part
we can argue that only two NS5-branes are needed in ten dimensions.

Recall that the curve had an additional symmetry $\sigma_3: t\to -t$.
Because this leaves the curve intact the effective period of $x_{10}$
is $2\pi \tilde R = \pi R$. Therefore instead of $t$ we should use
\[
 \tilde t =\mu\exp (-(x^6 + {\mathrm i} x^{10} )/\tilde R)= t^2
\] 
as the variable of the curve. This gives
\begin{equation}
{\tilde t} + \mu^2 /{\tilde t} + 2\mu - v^2 P_{2n} =0,
\label{NewSp}
\end{equation}
where the remaining symmetries $\sigma_1$ and $\sigma_2$ can be
realized as the action of an O6. Taking the quotient with respect to
these gives the genus $n$ curve:
\[
\C'_{{\rm Sp}(2n)}:\quad r^2 = \xi P_n(\xi) (\xi P_n(\xi) -4 \mu),
\]
where $r={\tilde t} - \mu^2 /{\tilde t}$. This curve  is
hyperelliptic, unlike the one we started from. 

Solving the Sp($2n$) curve (\ref{curves}) for $\tilde t$ we get,
\[
\tilde t= \half \left[ v^2 P_n(v^2) -2\mu  
\pm v \sqrt{P_{n}(v^2)(v^2 P_n(v^2) -4\mu)}\,\right] .
\] 
The point at $v=0$ is not a branch point. Even though there is only
one solution at that point, $\tilde t= -\mu$, going around it does not
produce a jump in the phase. This implies that there is no wrapping
around the $x_{10}$ direction and therefore no D4-brane.  At this
point the curve is singular since $\partial F(v,\tilde t)/\partial v
|_{v=0}=\partial F(v,\tilde t)/\partial \tilde t |_{\tilde t
=-\mu}=0$.  Thus, the two sheets will be connected at $v=0$ but this
does not correspond to a dynamical D4-brane but to a singularity of
the curve. But there are still $2n+1$ branch cuts. The one on the
middle (see Fig. \ref{Sp4fig}) comes from the factor $(v^2 P_n(v^2)
-2\mu)$ and will in the $\mu\to 0$ limit collapse to a D4-brane stuck
in the origin. This differs from the straightforward interpretation of
the zeroes of the polynomials as the positions of the branes, which
would give either $2n+2$ D4-branes, none of them at the origin, or
$2n$ branes plus two branes on the origin. For the latter case,
however, one needs to take a curve that does not come from the
integrable system \cite{BrSoThYa:SOSp}.

As for SO($2n+1$) we could have gotten the two-fold cover curve
(\ref{NewSp}) for Sp($2n$) directly from the M-theory curves
(\ref{M-curves}) by choosing a differently charged orientifold $2k=-4$
and another polynomial $x+y - P_{2n}(v^2)+2\mu v^{-2}=0$. But then $x$
and $y$ would scale like $ \Lambda^{2h^\vee_g-2}$ and we would need to
introduce the term $2\mu v^{-2}$ by hand. Therefore we think that the
Sp($2n$) curves in M-theory must be of second order in $x,y$ and
consequently of fourth order in $t$, but the corresponding
ten-dimensional brane diagram can be described with only two
NS5-branes.

%%%%%%%%%%%%%%%%%%%%%%%%%%%%%%%%%%%%%%%%%%%%%%%%%%%%%%%%%%%%%%%%%%%%
\section{Branes and the Curves for G$_2$ and E$_6$}
\label{G2}

Now we can proceed to wrap the M-theory five brane on the Toda curve
of exceptional groups and use the techniques developed in the previous
sections to investigate what happens in the ten dimensional limit.  We
will study G$_2$ in detail and after that make brief comments on
E$_6$.

\subsection{{\rm G}$_2$}

The G$_2$ curve is
\begin{equation}
\C_{{\rm G}_2}:
\quad 3[t+\mu/t]^2 + 2 v^2 A_2(v^2)[t+\mu/t]- v^2 B_6(v^2) -12\mu=0,
\label{g2curve}
\end{equation}
where $ A_2(v^2)= u_2 - 3 v^2$ and $B_6(v^2)= v^6 - 2 u_2 v^4 + u_2^2
v^2 - u_6$. It has genus 11 and it is of fourth order in $t$.  Unlike
for Sp($2n$) there apparently is no symmetry that would allow us to
find a two fold cover curve which is equivalent from the ten
dimensional point of view.  To see this, we first find the quotient
curve $\C'_{{\rm G}_2}$ that contains the Prym variety.

The obvious symmetries of $\C_{{\rm G}_2}$ are $\sigma_1$ and $\sigma_2$
in (\ref{sigma12}). We could obtain a double cover curve by taking the
quotient with respect to these, but the resulting curve has genus one
--- too small to contain the Prym. There is a third symmetry though:
\begin{equation}
\sigma_{\mathrm G}:
\quad t + \mu /t \to -[t + \mu /t + \frac{2}{3} v^2 A_2(v^2) ]. 
\label{sigmaG}
\end{equation}
Therefore, good variables are the ones invariant under the combined
action of $\sigma_1$, $\sigma_2$ and $\sigma_{\mathrm G}$:
\begin{equation}
\xi = v^2,\quad z=  v [t + \mu /t +  \frac{1}{3}v^2 A_2(v^2) ].
\label{z}
\end{equation}
They give a genus two curve \cite{Don:SW}
\begin{equation}
\C'_{{\rm G}_2}:\quad 9 z^2 - \xi\left(3 \xi B_3(\xi) + \xi^2 {A_1}^2
(\xi) +36 \mu \right) =0
\label{g2prym}
\end{equation}
which is hyperelliptic even when the original curve was
not\footnote{The remark made in \cite{Ito} on the possibility of
writing the Picard-Fuchs equations for G$_2$ with the help of elliptic
differential operators is no doubt a signal of this.}.  This curve is
not the same as the hyperelliptic curve proposed for G$_2$ in
\cite{AliArMan:g2hyper,DanSun:Ex}, whose physical validity was later
questioned in \cite{LanPieGid:G2}.  Furthermore, it can be checked
that the discriminant of the reduced curve $\Delta_{\C'}$ is the same
as the quantum discriminant $\Delta_q$ of G$_2$ field theory
calculated in \cite{LanPieGid:G2}:
\begin{eqnarray}
&\Delta_q &= \mu \Delta_{-}^2 \Delta_{+}^2\nonumber \\ 
&\Delta_{\pm}&
= -4 u_6 {u_2}^3 + 27 {u_6}^2 \mp 32 {u_2}^4 \sqrt{\mu} \pm 216 u_2
u_6 \sqrt{\mu} -144 {u_2}^2 \mu \mp 6912 \mu^{3/2} .
\end{eqnarray}  
It does differ from the discriminant of the original four-fold cover
curve $\C_{{\rm G}_2}$
\[
\Delta_{\C}=\mu ({u_6}^2 - 16 {u_2}^2\mu )\Delta_{-}^3 \Delta_{+}^3
\]
by a prefactor, which was argued in \cite{LanPieGid:G2} to correspond
to an unphysical singularity.

In the previous section we saw that for Sp($2n$) the original spectral
curve is also of fourth order in $t$ but it has a symmetry $t \to -t$
which allows us to define $\tilde t = t^2$. This is equal to a
rescaling of the period of the eleventh coordinate $x^{10}$ and thus
will not show in the ten dimensional brane diagram.  Clearly,
considering the symmetries of the G$_2$ curve, no scaling of $x^{10}$
will able us to find a two fold cover equivalent to (\ref{g2curve}).  So
if we want to construct a brane diagram for this theory it would have
to originate from the four fold cover curve $\C_{{\rm G}_2}$ and therefore
 have four NS5-branes.  We are going to investigate whether such a
configuration makes sense and if it does not, why. Also, we want to
understand how it would differ from the brane configurations for
product gauge groups.

The branch points of the curve are located at the zeroes of
\begin{equation}
P_8(v)= 3 B_6(v) v^2 + A_2(v)^2 v^4 + 36\mu =0
\label{shortsqrt}
\end{equation}
and $ v^2[3B_6(v)+  2 v^2 A_2(v)^2 \pm2  A_2(v) \sqrt{P_8(v)}\, ]=0$,
which is equivalent to
\begin{equation}
v^2 P_6^\pm(v) = v^2 [B_6(v) \pm 4 \sqrt \mu\, A_2(v)]=0.
\label{pmsqrt}
\end{equation}
Note that as for Sp($2n$) the $v=0$ is not a branch point but a
singularity of the curve.  The two sets of branch points have very
different characteristics.  The global behavior of the cuts is a
complicated issue.  Fortunately, for our purposes it suffices to study
some general features.

\begin{figure}
  \noindent
    \centering\includegraphics[width=10cm]{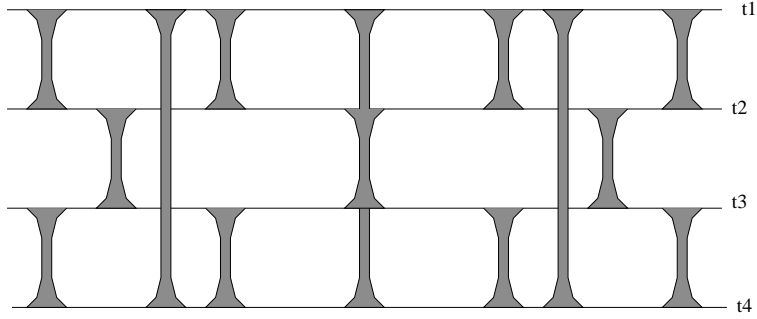}
    \caption{G$_2$ curve as a four fold cover of the $v$ plane.}
\label{coverfig}
\end{figure}

We now solve (\ref{g2curve}) for $t$ as a function of $v$ and label
the sheets $t_1,t_2,t_3$ and $t_4$. For any point $v_0$ satisfying
$P_8(v_0)=0$ we see that $t_1(v_0)=t_2(v_0) $ and $t_3(v_0)=t_4(v_0)
$. Therefore, for each pair of branch points coming from
(\ref{shortsqrt}) there will be one tube connecting the $t_1$ and
$t_2$ sheets and a second tube connecting the $t_3$ and $t_4$ sheets
at the same value of $v=v_0$, giving a total of eight cuts (see Figure
\ref{coverfig}).
The branch points satisfying (\ref{pmsqrt}) give rise to six branch
cuts that join either $t_1$ and $t_4$ ( $P_6^{-}(v)=0$) or $t_2$ and
$t_3$ ( $P_6^{+}(v)=0$ ). Unlike the previous case each cut
corresponds to only one tube joining a pair of sheets.

If we identify the cuts with D4-branes we immediately see why this
configuration is different from a product group: for product groups
the positions of branes between different pairs of NS5-branes are not
related. The D4-branes between the $k^{th}$ and the $(k+1)^{th}$
NS5-branes are free to move independently of the position of the
branes between the $(k+1)^{th}$ and the $(k+2)^{th}$ NS5-brane.  For
the G$_2$ four fold cover for any value of the moduli the branch cuts
between $t_1$ and $t_2$ have to be located at the same $(x^4,x^5)$
position as the cuts connecting $t_3$ and $t_4$. A similar thing
happens for the brane configuration of SO($2n$) with matter in the
symmetric representation \cite{LanLop:New}, as a result of the
orientifold six-plane in the middle. Moreover, because G$_2$ has only
two moduli the positions of $P_8$-cuts depend in some way on the
positions of the cuts that originate from $P_6^\pm$. Even classically,
if we follow the prescription of identifying the number of branes with
zeroes of the polynomials, we run into trouble. We would obtain a
configuration with four NS5-branes, three D4-branes in the outermost
columns and eight in the middle one. Seemingly, this describes a
theory with ${\rm Sp}(2) \times {\rm SO}(8) \times {\rm Sp}(2)$ gauge
group, but the curve only contains two moduli and one scale, $\mu$,
and therefore can not describe a product gauge group.

A serious complication is that not all the cuts collapse to points
when we go to ten dimensions, which was the requirement of having a
well defined classical ten dimensional limit. All the cuts coming from
the polynomials $P^\pm_6$ (\ref{pmsqrt}) and one cut from
(\ref{shortsqrt}) have the desired behaviour when $\mu\to 0$, but the
remaining three cuts do not.  Therefore it seems that starting from
the curve (\ref{g2curve}) it is not possible to get a ten dimensional
brane configuration with G$_2$ field theory on the world volume.

If we try to construct a curve such that at least the discriminant
would have same zeroes as the discriminant of the G$_2$ field theory,
and also would have a well defined ten dimensional limit we will also
encounter problems.  For example, consider
\begin{equation}
 2 A_2(v^2) v^2 (t+\mu/t)- B_6(v^2) =0,
\label{almostG2}
\end{equation}
which amounts to sending the outermost NS5-branes to infinity.  The
branch points are at $P_6^- P_6^+ =0$ and the discriminant of the
curve is $ ({u_6}^2 - 16 {u_2}^2\mu )\Delta_{-} \Delta_{+}$. Now the
cuts will collapse to points, creating a brane configuration with two
NS5-branes and six four branes between them, four semi-infinite
D4-branes on each side (two of them located at $v=0$) and an O6 in the
middle.  The semi-infinite branes do not generate hypermultiplets
since there are no free moduli associated to any of them. The
discriminant is the correct G$_2$ discriminant, modulo an unphysical
factor, but we cannot obtain the G$_2$ Prym from it.  The Jacobians of
the new curve and $\C_{{\rm G}_2}$ can not be the the same because
there is no way of getting the correct genus two curve from
(\ref{almostG2}) by taking the quotient with respect to the symmetries
--- $\C'_{{\rm G}_2}$ in (\ref{g2prym}) depends only on the polynomial
$P_8(v)$, not on $P_6^\pm$. Therefore, even if this brane
configuration does have some of the characteristics of a G$_2$ it can
not contain all the information needed. We interpret these findings as
an indication that it is not possible to have a weakly coupled Type
IIA brane configuration that will reproduce a G$_2$ gauge theory.

Nevertheless, in the spirit of our previous examples we would like to
implement the symmetries directly in M-theory to isolate the
physically significant part of the curve. In order to get the correct
gauge groups in the D4 world-volume gauge theory one has to project
out part of the Chan-Paton degrees of freedom. For all other groups we
have studied so far these projections were equivalent to modding out
the symmetries of the curve, to find the genus $n$ part that contains
the Prym. For G$_2$ the symmetries $\sigma_1$ and $\sigma_2$ can be
realized as an O6, but since this gives a curve with genus one instead
of two, it is not enough. We would need to implement also the new
symmetry (\ref{sigmaG}).

The transformation $\sigma_{\mathrm G}$ does square to identity, but
it does not admit an interpretation as an orientifold plane in terms
of the coordinates $t$ and $v$. In order to realize this symmetry we
would need not an orientifold plane but a more complicated object in
M-theory, which would live in the invariant locus $t+ \mu /t +
\frac{1}{3} v^2 A_2(v)=0$ of the transformation $\sigma_{\rm G}$. This
seems to be a kind of curved orientifold surface.  After a coordinate
transformation to $z$ defined in (\ref{z}) $\sigma_{\rm G}$ reads
$\frac{z}{v} \to -\frac{z}{v}$, which is the symmetry of an ordinary
orientifold plane. However, after this transformation the O6 plane
which corresponds to symmetries $\sigma_1, \sigma_2$ will turn into a
curved surface.  It would be interesting to see if it is possible to
construct a M-theory background, like the Atiyah-Hitchin space for O6,
that would realize all the symmetries $\sigma_2,\sigma_2$ and
$\sigma_{\rm G}$ simultaneously.

\subsection{{\rm E}$_6$}

The E$_6$ curve presents many of the characteristics seen in G$_2$.
It is a genus 34 curve that can be realized as a four-fold cover of
the $v$-plane (\ref{curves}).  The branch cuts of the curve have the same
behavior as in the G$_2$ case. Namely, they do not collapse to points
when compactifying ten dimensions and thus will not describe D4-branes
in this limit.

One of our original motivations was to examine the breaking of
exceptional groups to get matter in the spinor representation. It is
perfectly possible to take a curve, write the invariants $u_i$ of the
bigger group in terms of the moduli of the group that it breaks into
(for explicit construction for $E_6\to SO(10)\times U(1)$ see
\cite{LerWar:LG}) and recover in some limit of the moduli space a new
curve which describes a group with a specific matter content.
Unfortunately, as we have argued, there is no guarantee that the
resulting curve will have a good limit in weakly coupled Type IIA
theory.
 
For E$_6$ it is not known if a genus six curve that contains the
Prym-Tjurin \cite{Don:SW,Kanev} variety can be constructed explicitly.
Thus, we can only speculate that even more complicated object than a
curved orientifold would be needed in M-theory in order to obtain this
subvariety.

%%%%%%%%%%%%%%%%%%%%%%%%%%%%%%%%%%%%%%%%%%%%%%%%%%%%%%%%%%%%%%%%%%%%

\section{Conclusions}
\label{conclusions}

We have seen that in the brane configurations for classical groups we
can generically identify D4-branes with branch cuts of the
Seiberg-Witten curve $\Sigma$.  The need to introduce O6 or O4-planes
to the brane configuration can be understood as the symmetries that
have to be quotioned out in order to obtain the physically significant
curve that has the Prym variety as its Jacobian.

We found several ways how the G$_2$ Toda curve differs from those of
the classical groups. It does not have a well defined classical ten
dimensional limit that would reduce the cuts to D4-branes.  We think
it is not possible to find a brane configuration in weakly coupled
Type IIA string theory which would describe a G$_2$ field theory on
the D4 world-volume. The situation for E$_6$ is similar and even more
complicated.

Moreover, the symmetries of the G$_2$ curve can not be realized as
orientifold planes. Instead we would need, in addition to an O6 plane,
a curved orientifold surface.  It would be interesting to see if it is
possible to construct an M-theory background that would realize all
the symmetries of G$_2$ simultaneously.

\subsection*{Acknowledgements}

We would like to thank Eric D'Hoker, Jacques Distler and Sergey Cherkis
 for useful conversations and comments. 
E.C. thanks the Theory Group of UT  Austin and 
P.P. the TEP Group in UCLA for their 
hospitality. The work of P.P. is supported by NSF grant PHY-9511632
and the Robert A. Welch Foundation and the work of E.C. by NSF PHY-9531023. 

%\clearpage
%%%%%%%%%%%%%%%%%%%%%%%%%%%%%%%%%%%%%%%%%%%%%%%%%%%%%%%%%%%%%%%%

%%%%%%%%%%%%%%%%%%%%%%%%%%%%%%%%%%%%%%%%%%%%%%%%%%%%%%%%%%%%%%%%


\baselineskip 0.4cm
\begin{thebibliography}{99}

\bibitem{HanWit:3d}
A. Hanany and E. Witten,
{\em Type IIB Superstrings, BPS Monopoles, 
and Three-Dimensional Gauge Dynamics},
\np{492}{1997}{152},
 {\tt hep-th}/9611230.

\bibitem{Wit:M}
E. Witten,
{\em Solutions of Four-Dimensional 
Field Theories Via M Theory},
\np{500}{1997}{3}, 
{\tt hep-th}/9703166


\bibitem{Bar:Rot}
J. L. F. Barb\'on,
{\em Rotated Branes and $N=1$ Duality}, 
\pl{402}{1997}{59}, 
{\tt hep-th}/9703051

\bibitem{Wit:BraQCD}
E. Witten,
{\em Branes and the Dynamics of QCD},
\np{507}{1997}{658},
{\tt hep-th}/9706109

\bibitem{EliGiKuRaSch}
S. Elitzur, A. Giveon, D. Kutasov, E. Rabinovici, A. Schwimmer,
{\em Brane Dynamics and N=1 Supersymmetric Gauge Theory },
\np{505}{1997}{202}, 
{\tt hep-th}/9704104

\bibitem{GivKut:Bra}
 A. Giveon and D. Kutasov,
{\em   Brane Dynamics and Gauge Theory },
{\tt hep-th}/9802067

\bibitem{LanLopLow:Ori}
K. Landsteiner, E. Lopez and D.A. Lowe,
{\em N=2 Supersymmetric Gauge Theories, Branes and Orientifolds},
\np{507}{1997}{197}, 
{\tt hep-th}/9705199

\bibitem{BrSoThYa:SOSp}
A. Brandhuber, J. Sonnenschein, S.~Theisen and S.~`Yankielowicz,
{\em M Theory and Seiberg-Witten 
Curves: Orthogonal and Symplectic Groups},
\np{504}{1997}{175}, 
{\tt hep-th}/9705232 

\bibitem{LanLop:New}
K. Landsteiner and E. Lopez,
{\em New Curves from Branes},
\np{516}{1998}{273},
{\tt hep-th}/9708118

\bibitem{DaMu:F}
K. Dasgupta, S. Mukhi,
{\em F-theory at constant coupling},
\pl{B385}{1996}{125},
{\tt hep-th}/9606044

\bibitem{Jo}
A. Johansen,
{\em A comment on BPS states in F-theory in 8 dimensions},
\pl{B395}{1997}{36},
{\tt hep-th}/9608186

\bibitem{GaZw:excep}
M. Gaberdiel, B. Zwiebach,
{\em Exceptional Groups from Open Strings},
\np{518}{1998}{151},
{\tt hep-th}/970913

\bibitem{GaHaZw:tensor1}
M. Gaberdiel, B. Zwiebach,
{\em Tensor Constructions of Open String Theories 1: Foundations},
\np{505}{1997}{569},
{\tt hep-th}/9705038

\bibitem{MarWar:Int}
E. Martinec and N. P. Warner,
{\em Integrable Systems and Supersymmetric Gauge Theory},
\np459{}{1996}{97},
{\tt hep-th}/9709161 

\bibitem{LerWar:E6}
W. Lerche and N. P. Warner,
{\em Exceptional SW Geometry from ALE Fibrations}, 
\pl{423}{1998}{79},
{\tt hep-th}/9608183

\bibitem{Poul:dual}
P. Pouliot,
{\em Chiral Duals of Non-Chiral SUSY Gauge Theories},
\pl{359B}{1995}{108},
{\tt hep-th}/9507018

\bibitem{BeChoKrStr:so10}
M. Berkooz, P. Cho, P. Kraus, M.J. Strassler,
{\em Dual Descriptions of SO(10) SUSY Gauge Theories with Arbitrary Numbers 
of Spinors and Vectors},
\prd{D56}{1997}{7166},
{\tt hep-th}/9705003

\bibitem{KlLeMaVaWa}
A. Klemm, W. Lerche, P. Mayr, C. Vafa, N. Warner,
{\em Self-dual Strings and N=2 Supersymmetric Field Theory},
\np{477}{1996}{746},
{\tt hep-th}/9604034

\bibitem{ItoMor:intSW}
H. Itoyama and A. Morozov
{\em Integrability and Seiberg-Witten theory: Curves and Periods
      Authors: H. Itoyama, A. Morozov},
\np{477}{1996}{855}, 
{\tt hep-th}/9511126
  
\bibitem{GKMMM}
A.Gorsky, I.Krichever, A.Marshakov, A.Mironov, A.Morozov,
{\em Integrability and Seiberg-Witten Exact Solution },
\pl{355}{1995}{466}, 
{\tt hep-th}/9505035

\bibitem{MarWar:proof}
E. J. Martinec and N. P. Warner,
{\em Integrability and N=2 Gauge Theory : A Proof}, 
{\tt hep-th}/9511052

\bibitem{DHokPho:CalMos}
E. D'Hoker and D.H. Phong,
{\em Calogero-Moser Systems in $SU(N)$ Seiberg-Witten Theory},
\np{513}{1998}{405},
 {\tt hep-th}/9709053

\bibitem{DonWit:YMInt}
 R. Donagi and E. Witten,
{\em Supersymmetric Yang-Mills Systems And Integrable Systems},
\np{460}{1996}{299},
{\tt hep-th}/9510101

\bibitem{AhnOhTatar:sp}
C. Ahn, K. Oh, R. Tatar
{\em Sp(N) Gauge Theories and M Theory Fivebrane},
{\tt hep-th}/9708127

\bibitem{AhnOhTatar:so}
C. Ahn, K. Oh, R. Tatar,
{\em M Theory Interpretation for Strong Coupling Dynamics of SO(N) Gauge 
Theories},
\pl{416}{1998}{75},
{\tt hep-th}/9709096

\bibitem{AhnOhTatar:o6}
C. Ahn, K. Oh and R. Tatar,
{\em Comments on SO/SP Gauge Theories from Brane Configurations with an O(6)
Plane},
{\tt hep-th}/9803197

\bibitem{ErlNaqRan:Prod}
J. Erlich, A. Naqvi and L. Randall,
{\em The Coulomb Branch of N=2 Supersymmetric
Product Group Theories from Branes},
{\tt hep-th}/9801108

\bibitem{Don:SW}
 R. Donagi, 
{\em Seiberg-Witten integrable systems},
alg-geom/9705010

\bibitem{EvaJonSha}
N. Evans, C.V. Johnson and A.D. Shapere,
{\em Orientifolds, Branes, and Duality of 4D Gauge Theories},
\np{505}{1997}{251}, 
{\tt hep-th}/9703210

\bibitem{Seib:IRDyn}
N. Seiberg,
{\em IR Dynamics on Branes and Space-Time Geometry },
\pl{384}{1996}{81},
 {\tt hep-th}/9606017

\bibitem{Sen}
A. Sen,
{\em A Note on Enhanced Gauge Symmetries in M- and String Theory},
J.High Energy Phys.{\bf 9} (1997) 001, {\tt hep-th}/9707123

\bibitem{SeiWit:Comp3d}
N. Seiberg and E. Witten,
{\em Gauge Dynamics and Compactification to Three Dimensions}, 
{\tt hep-th}/9607163

\bibitem{AliArMan:g2hyper}
M. Alishahiha, F. Ardalan, F. Mansouri,
{\em The Moduli Space of the Supersymmetric G(2) Yang-Mills Theory },
\pl{381}{1996}{446},
{\tt hep-th}/9512005

\bibitem{DanSun:Ex}
 U. H. Danielsson and B. Sundborg,
{\em Exceptional Equivalences in N=2 Supersymmetric Yang-Mills Theory},
\pl{370}{1996}{83},
{\tt hep-th}/9511180

\bibitem{LanPieGid:G2}
 K. Landsteiner, J. M. Pierre and  S. B. Giddings,
{\em On the Moduli Space of N = 2 Supersymmetric $G_2$ Gauge Theory },
\prd{55}{1997}{2367},
{\tt hep-th}/9609059

\bibitem{Ito}
K. Ito,
{\em Picard-Fuchs Equations and Prepotential in N=2 Supersymmetric $G_2$
Yang-Mills Theory },
\pl{406}{1997}{54},
{\tt hep-th}/9703180

\bibitem{LerWar:LG} 
W. Lerche and N. P. Warner,
{\em Polytopes and Solitons in Integrable N=2 Supersymmetric
Landau-Ginzburg Theories}, \np{358}{1991}{571}

\bibitem{Kanev}V. Kanev,
{\em Spectral Curves, Simple Lie Algebras and Prym-Tjurin Varieties },
Proceedings of Symposia in Pure Mathematics, Vol. {\bf 49} (1989) 627 


\end{thebibliography}
\end{document}